\documentclass[11pt]{article}

\usepackage{latexsym,color,amsmath,amssymb}

\def\marrow{\marginpar[\hfill$\longrightarrow$]{$\longleftarrow$}}
\def\micha#1{\textsc{(Micha says: }\marrow\textsf{#1})}

\newtheorem{theorem}{Theorem}

\newtheorem{proposition}[theorem]{Proposition}

\def\reals{{\mathbb R}}
\def\eps{{\varepsilon}}
\def\CP{{\mathbb CP}}
\def\ee{{\bf e}}
\def\nn{{\bf n}}
\def\xx{{\bf x}}
\def\xd{{\bf x}'}
\def\xdd{{\bf x}''}

\begin{document}

\title{On Lines and Joints\thanks{%
  Work on this paper has been partly supported by the
  Hermann Minkowski--MINERVA Center for Geometry at Tel Aviv
  University.
  Work by Micha Sharir has also been supported
  by NSF Grants CCF-05-14079 and CCF-08-30272,
  by a grant from the U.S.-Israeli Binational Science Foundation,
  and by grant 155/05 from the Israel Science Fund.
  Work by Haim Kaplan has also been supported by Grant 975/06 from the
  Israel Science Fund, and  the United states - Israel
Binational Science Foundation, project number 2006204.} }

\author{
Haim Kaplan\thanks{%
School of Computer Science, Tel Aviv University,
Tel Aviv 69978 Israel;
\textsl{haimk@post.tau.ac.il}.}
\and
Micha Sharir\thanks{%
School of Computer Science, Tel Aviv University,
Tel Aviv 69978 Israel and Courant Institute of Mathematical Sciences,
New York University, New York, NY 10012, USA;
\textsl{michas@post.tau.ac.il}.}
\and
Eugenii Shustin\thanks{%
School of Mathematical Sciences, Tel Aviv University,
Tel Aviv 69978 Israel;
\textsl{shustin@post.tau.ac.il}.}
}

\maketitle

\thispagestyle{empty}


\begin{abstract}
Let $L$ be a set of $n$ lines in $\reals^d$, for $d\ge 3$.
A {\em joint} of $L$ is a point incident to at least $d$ lines of $L$,
not all in a common hyperplane.
Using a very simple algebraic proof technique, we show that the
maximum possible number of joints of $L$ is $\Theta(n^{d/(d-1)})$.
For $d=3$, this is a considerable simplification of the orignal
algebraic proof of Guth and Katz~\cite{GK}, and of
the follow-up simpler proof of Elekes et al.~\cite{EKS}.
\end{abstract}

Let $L$ be a set of $n$ lines in $\reals^d$, for $d\ge 3$.
A {\em joint} of $L$ is a point incident to at least $d$ lines of $L$,
not all in a common hyperplane.

A simple construction, using the axis-parallel
lines in a $k\times k\times \cdots \times k$ grid, for
$k=\Theta(n^{1/(d-1)})$, has
$dk^{d-1}=\Theta(n)$ lines and $k^d=\Theta(n^{d/(d-1)})$ joints.

In this paper we prove that this is a general upper bound. That is:
\begin{theorem} \label{main}
The maximum possible number of joints in a set of $n$ lines in
$\reals^d$ is $\Theta(n^{d/(d-1)})$.
\end{theorem}

\paragraph{Background.}
The problem of bounding the number of joints, for the
$3$-dimensional case, has been around for almost 20 years
\cite{CEG+,FS,Sh:j} (see also \cite[Chapter 7.1, Problem 4]{BMP05}),
and, until very recently, the best known upper
bound, established by Sharir and Feldman~\cite{FS}, was
$O(n^{1.6232})$.  The proof techniques were rather complicated,
involving a battery of tools from combinatorial geometry, including
forbidden subgraphs in extremal graph theory, space decomposition
techniques, and some basic results in the geometry of lines in space
(e.g., Pl\"ucker coordinates).

Wolff \cite{Wol96} observed a connection between the problem of
counting joints to the Kakeya problem. Bennett et al.~\cite{BCT05}
exploited this connection and proved an upper bound on the number
of so-called {\em $\theta$-transverse} joints in $\reals^3$, namely,
joints incident to at least one triple of lines for which the volume
of the parallelepiped generated by the three unit vectors along these
lines is at least $\theta$. This bound is
$O(n^{3/2+\eps}/\theta^{1/2+\eps})$, for any $\eps>0$, where the
constant of proportionality depends on $\eps$.

It has long been conjectured that the correct upper bound
on the number of joints (in three dimensions) is $O(n^{3/2})$,
matching the lower bound just noted. In a rather dramatic recent
development, Guth and Katz \cite{GK} have settled the conjecture in
the affirmative, showing that the number of joints (in three
dimensions) is indeed $O(n^{3/2})$.
Their proof technique is completely different, and
uses fairly simple tools from algebraic geometry. In a follow-up
paper by Elekes et al.~\cite{EKS}, the proof has been further
simplified, and extended (a) to obtain bounds on the number of
incidences between $n$ lines and (some of) their joints, and
(b) to handle also {\em flat} points, which are points incident to
at least three lines, all coplanar.

As far as we know, the problem has not yet been studied for $d > 3$.

In this paper we give a very simple and short proof of
Theorem~\ref{main}; that is, we obtain a tight bound for the
maximum possible number of joints in any dimension. The proof uses
an algebraic approach similar to that of the other proofs, but is
much simpler, shorter and direct.

We note that this paper does not subsume the previous paper
\cite{EKS}, because the new proof technique cannot handle the problem
of counting incidences between lines and joints, nor can it handle
flat points. Nevertheless, it is our hope that these extensions
would also be amenable to similarly simpler proof techniques.

\paragraph{Analysis.}
We will need the following well-known result from algebraic
geometry; see proofs for the 3-dimensional case in \cite{EKS,GK}.
We include the easy general proof for the sake of completeness.
\begin{proposition} \label{prop4}
Given a set $S$ of $m$ points in $d$-space, there exists a
nontrivial $d$-variate polynomial $p(x_1,\ldots,x_d)$ which
vanishes at all the points of $S$, whose degree is at most
the smallest integer $b$ satisfying ${b+d\choose d} > m$.
\end{proposition}
\noindent{\bf Proof.}
A $d$-variate polynomial of degree $b$ has
${b+d\choose d}$ monomials, and requiring it to vanish at
$m < {b+d\choose d}$ points yields $m$ linear homogeneous
equations in the coefficients of these monomials.  Such an
underdetermined system always has a nontrivial solution.
$\Box$

\bigskip

\noindent{\bf Proof of Theorem~\ref{main}.}
We only need to prove the upper bound. Let $L$ be a set of $n$ lines
in $\reals^d$, and let $J$ denote the set of joints of $L$. Put
$m=|J|$, and assume to the contrary that $m>An^{d/(d-1)}$, for some
constant parameter $A$, depending on $d$, which we will fix shortly.

\paragraph{Pruning.}
We first apply the following iterative pruning process to $L$.
As long as there exists a line $\ell\in L$ incident to fewer than
$m/(2n)$ points of $J$, we remove $\ell$ from $L$, remove its
incident points from $J$, and repeat this step with respect to
the reduced sets of lines and points (keeping the threshold
$m/(2n)$ fixed).  In this process we delete at most $m/2$ points.
We are thus left with a subset of the original lines, each
incident to at least $m/(2n)$ surviving points, and each surviving
point is a joint in the set of surviving lines, that is, it is
incident to at least $d$ surviving lines, not all in a common
hyperplane.  For simplicity, continue to denote these sets
as $L$ and $J$.

\paragraph{Vanishing.}
Applying Proposition~\ref{prop4}, we obtain a nontrivial
$d$-variate polynomial $p$ which vanishes at all the (at most)
$m$ points of $J$, whose degree is at most the smallest integer
$b$ satisfying ${b+d\choose d} \ge m+1$, so the degree is at most
$$
b \le \lceil (d!m)^{1/d} \rceil \le 2(d!m)^{1/d} .
$$

We choose $A$ so that the number of points on each (surviving) line is
greater than $b$. That is, we require that $m/(2n) > 2(d!m)^{1/d}$,
or that $m > \left(2^{d+1}d!\right)^{1/(d-1)} n$, which will hold
if we choose $A > (2^{d+1}d!)^{1/(d-1)}$.
(Asymptotically, the right-hand side
is only slightly larger than $d$.)

With this choice of $A$, the polynomial $p$ vanishes on at least
$m/(2n) > b$ points on each line in $L$. Hence, $p$ vanishes
identically on every line of $L$.

\paragraph{Differentiating.}
Fix a point $a\in J$, and let $\ell$ be a line of $L$ incident to $a$.
Parametrize points on $\ell$ as $a+tv$, where $v$ is a (unit)
vector in the direction of $\ell$. We have, for $t$ sufficiently
small,
$$
p(a+tv) = p(a) + t \nabla p(a)\cdot v + O(t^2) .
$$
Since $p\equiv 0$ on $\ell$, we must have $\nabla p(a) \cdot v = 0$.
This holds for every line of $L$ incident to $a$. But since $a$ is a
joint, the directions of these lines span the entire $d$-space, so
$\nabla p(a)$, being orthogonal to all of them,
must be the zero vector. That is, all the first-order
derivatives of $p$ vanish at $a$.

Consider one of these derivatives, say $p_{x_1}$, whose degree is at
most $b-1$, and consider a line $\ell\in L$.
Since $\ell$ contains more than $b-1$ points of $J$, and $p_{x_1}$
vanishes at each of these points, $p_{x_1}$ must vanish identically
on $\ell$. That is, all first-order derivatives of $p$ vanish
on all the lines of $L$.

We now repeat the above analysis to each of these derivatives.
Consider again $p_{x_1}$, say, and pick a point $a\in J$ and an
incident line $\ell\in L$. Using the above notation, we have, in
complete analogy,
$$
p_{x_1}(a+tv) = p_{x_1}(a) + t \nabla p_{x_1}(a)\cdot v + O(t^2) ,
$$
for $t$ small enough, implying that
$\nabla p_{x_1}(a)$ is orthogonal to each of the lines incident to
$a$, and therefore must be zero. Applying this argument to each of
the first-order derivatives, we conclude that all second-order
derivatives of $p$ vanish at every point of $J$. Since the degree
of these derivatives is at most $b-2$, they all vanish identically on
every line of $L$.

Iterating this process, we conclude that {\em all} partial
derivatives of $p$ vanish identically on all the lines of $L$. This
however is impossible, because eventually we reach derivatives which
are nonzero constants on $\reals^d$.
This completes the proof of the theorem.
$\Box$

\paragraph{Extensions.}
{\bf (1)} Theorem~\ref{main} extends in a straightforward manner to
the case of joints generated by algebraic curves in $\reals^d$. Let
$F_i(t) = (x^{(i)}_1(t),x^{(i)}_2(t),\ldots,x^{(i)}_d(t))$ be an
algebraic curve of degree $n_i$, and put $n = \sum_i n_i$. Define a
joint to be a point $a$ incident to at least $d$ curves, such that
the tangents at $a$ to the curves incident to $a$ are not all 
in a common hyperplane. Then the number $m$ of
joints is at most $An^{d/(d-1)}$, with the same constant $A$ as in
the case of lines. Indeed, in an initial pruning step we get rid of
those curves $F_i$ which pass through fewer than $mn_i/(2n)$ joints,
and of the joints lying on these curves. We lose in this pruning at
most $m/2$ joints. We then construct, as before, a polynomial $p$,
of degree $b\approx m^{1/d}$, which vanishes on all the surviving
joints. Each of the surviving curves $F_j$ contains at least
$mn_j/(2n)$ surviving joints, and this quantity is larger than
$bn_j$, by the initial assumption in the proof. Since $p$ vanishes
at all these points, it follows from B\'ezout's theorem~\cite{CLO1}
that $p\equiv 0$ on each $F_j$. The proof then continues exactly as
before, showing that all partial derivatives of $p$, of any order,
vanish on all the surviving curves $F_j$, which yields a
contradiction, as above.

\noindent{\bf (2)} One can also derive the following variant of
Theorem~\ref{main}: Let $L$ be a set of $n$ lines in $\reals^d$, and
let $2\le s<d$ be an integer. Call a point $a$ an {\em $s$-joint} if
the affine hull of the directions of the lines of $L$ incident to
$a$ is (at least) $s$-dimensional. Then the number of $s$-joints of
$L$ is at most $A_sn^{s/(s-1)}$, where $A_s$ is the constant in
Theorem~\ref{main} for dimension $s$. For a proof, simply project
$L$ and the $s$-joints onto some generic $s$-flat ($s$-dimensional
affine subspace), and apply
Theorem~\ref{main} to the projected lines and points.

However, as shown in \cite{EKS} for the 3-dimensional case, the
{\em number} of lines of $L$ incident to a point is also an
important parameter, and not just the dimension of the flat
that they span. In analogy with the analysis in \cite{EKS}, one could
define a {\em $(k,s)$-joint} of $L$ to be a point $a$ incident to at
least $k$ lines of $L$, which span a flat of dimension at least $s$,
and ask for a bound on the number of $(k,s)$-joints, under some
further restrictions on $L$. For example, it is shown in \cite{EKS}
that the number of $(3,2)$-joints in $\reals^3$ is $O(n^{3/2})$,
provided that no plane contains more than $O(n)$ $(3,2)$-joints.

\paragraph{(3) Singular points of spatial algebraic curves.}
We finally state the following grand generalization of the above
results. Let $C$ be a reduced algebraic curve (i.e., a curve with no
multiple components) in the complex projective space $\CP^d$, for
$d\ge 3$. The {\em embedding dimension} of $C$ at a point $a\in C$
is the smallest integer $s\ge 1$ such that there exists an analytic
diffeomorphism $\varphi:U\to V$ of neighborhoods $U$ and $V$ of $a$
in $\CP^d$, so that $\varphi(C\cap U)$ is contained in an 
$s$-dimensional subspace of $\CP^d$ 
(cf.~\cite[Definition I.1.19 and Lemma I.1.24]{GLS}).
Clearly, the embedding dimension of a
curve $C$ at any of its non-singular points is $1$, and the
embedding dimension of a line configuration or of a configuration of
curves, as above, at a joint (respectively, at an $s$-joint) is $d$
(resp., $s$). We then have:

\begin{theorem} \label{alg_curves}
The number of points, where a reduced algebraic curve
$C\subset\CP^d$ of degree $n$ has embedding dimension $s$
($s\ge 2$) does not exceed $A(s) n^{s/(s-1)}$, where $A(s)$ is
the constant from the proof of Theorem~\ref{main}.
\end{theorem}

Indeed, the case $s=d$ (generalized joints) is treated precisely as
in the proof of Theorem \ref{main}, since a hypersurface containing
the curve $C$ must be singular at its generalized joints. In the
case $s<d$ we apply a generic projection onto $\CP^s$, similar to the
argument given above.

\paragraph{Acknowledgments.}
The authors thank Shakhar Smorodinsky for helpful discussions on this
problem. In particular, he observed independently that the analysis
also carries over to algebraic curves.

\end{document}